# Reversible Switching of Charge Injection Barriers at Metal/Organic-Semiconductor Contacts Modified with Structurally Disordered Molecular Monolayers

Ryo Nouchi,[1,2,3,a)] Masanori Shigeno,[4] Nao Yamada,[3] Tomoaki Nishino,[1] Katsumi Tanigaki[2,3] and Masahiko Yamaguchi[2,4,a)]

[1]*Nanoscience and Nanotechnology Research Center, Osaka Prefecture University, Sakai 599-8570, Japan*

[2]*WPI-Advanced Institute for Materials Research, Tohoku University, Sendai 980-8577, Japan*

[3]*Department of Physics, Graduate School of Science, Tohoku University, Sendai 980-8578, Japan*

[4]*Department of Organic Chemistry, Graduate School of Pharmaceutical Sciences, Tohoku University, Sendai 980-8578, Japan*

a) Electronic addresses: r-nouchi@21c.osakafu-u.ac.jp and yama@m.tohoku.ac.jp

**ABSTRACT**: Metal/semiconductor interfaces govern the operation of semiconductor devices through the formation of charge injection barriers that can be controlled by tuning the metal work function. However, the controlling ability is typically limited to being static. We show that a dynamic nature can be imparted to the interfaces using electrode surface modification with a structurally disordered molecular monolayer. The barrier height at the interfaces is altered significantly in a reversible way by an external electric field. As a result, a dramatic change in the carrier transport properties through the interfaces is observed, such as a reversible polarity reversal of metal/organic-semiconductor/metal diodes.

A self-assembled monolayer (SAM) is a monolayer-thick ordered film formed by the self-assembly of SAM constituents, such as organic molecules, through interconstituent interactions.[1] Due to the simple formation process (generally only immersion in a SAM-constituent solution), an extremely thin body (one monolayer that preserves the effective charge injection through the SAMs) and molecular diversity of the constituents (molecular synthesis techniques that introduce various substituent groups), SAMs are known to be efficient tuners of metal work functions (WFs)[2-5] and have been widely used to modify metallic electrodes of electronic devices, such as diodes[6-8] and field-effect transistors (FETs).[9,10] Figure 1 depicts two main mechanisms for the change of the WF by the modification of metal surfaces with self-assembled molecular monolayers. As a starting point, a bare metal surface is first illustrated in Fig. 1a. Electrons are known to spill out from the metal surface and the electron density becomes finite, even outside of the metal.[11] The resultant electric double layer at the metal surface forms an energy barrier for electrons inside the metal, which contributes to the metal WF. From a simple analogy, the introduction of additional electric double layers to the metal surface can alter the WF. Thus, if the molecules that form the monolayer possess a permanent electric dipole, the metal WF is changed as a function of the strength/direction of the dipole.[5-8] The metal WF increases (decreases) by the addition of dipoles with the same (opposite) direction as the inherent electric double layer at the bare metal surface, as schematically shown in Fig. 1b. Another effect of SAM formation should be additionally considered: the push-back (or "pillow") effect[12] that occurs at SAM-covered metal surfaces. The electrons spilled at the metal surfaces are pushed back inside the metal due to Pauli repulsion from the electron clouds of SAM molecules. Consequently, the strength of the surface electric double layer is weakened, which leads to a decrease in the WF (Fig. 1c).

The modification of metal surfaces with molecular monolayers enables the metal WF to be effectively tuned through the dipole and push-back effects. However, metal surfaces after modification are characterized by a single value of WF, which indicates that the surfaces possess no further tunability. If the WF is tunable, even after surface modification with a molecular monolayer, then the energy mismatch between the Fermi level of a metallic electrode and the semiconductor energy band can be resolved on demand, and thus very efficient charge carrier injection from a metallic electrode into a semiconductor layer can be achieved. In addition, post-modification tunability of the metal WF should offer the possibility to produce functional devices that exploit the switchable/reconfigurable characteristics.

The reason why modified metal surfaces do not generally have further tunability is due to the structural rigidity of the molecular monolayers[13]. Several efforts have been made to break the rigidity, so that surfaces modified with monolayers possess switchability. SAM structures are stabilized by interconstituent interactions, such as van der Waals and π-π interactions. Therefore, efforts have been largely devoted to produce low-density SAMs, including strategies that employ

SAM molecules with bulky terminal groups[14] and that decrease the SAM formation time.[15] The low density indicates a longer interconstituent distance, which weakens the interactions and thus leads to a reduction in the rigidity of the molecular monolayer. The former strategy[14] employs a multi-step process that starts with the formation of a molecular SAM with a bulky terminal group and is followed by removal of the bulky group by acid treatment, which leaves a SAM that exhibits low-density packing; reversible switching of the wettability of a Au(111) surface was achieved using this method. The latter strategy[15] simply decreased the SAM formation time to reduce the average packing density of the SAM molecules; however, the SAM formation process is rapid and many closely packed domains are formed even during the very early stage of the formation process,[16] which leads to low controllability of the SAM packing density. In this case, only electric current variations were evident in organic FETs, which were explained by a change in the charge injection efficiency due to a structural change of the SAMs on the electrodes.

In this Letter, we propose a simple single-step formation of a non-rigid molecular monolayer using a molecule with two bonding groups. A closely packed, well-ordered SAM structure on a certain surface is considered to be formed by the adsorption and subsequent lateral diffusion of the SAM constituents on the surface.[17] Therefore, if the binding energy of the constituent to the surface is too strong to allow the SAM constituents to diffuse freely at the surface, then an ordered SAM structure is not formed. Multiple bonds increase the binding energy significantly, which is known to generally result in a disordered monolayer (DM).[18] The intermolecular interaction should be weak in the DM, resulting in a non-rigid molecular monolayer (Fig. 1d). We synthesized a helicene derivative[19,20] with two thiol groups, 1,12-dimethyl-5,8-[4]helicenedithiol, which makes a strong covalent attachment to Au surfaces, and confirmed by scanning tunnelling microscopy (STM) that the helicene derivative indeed forms a DM on Au(111) surfaces.[21] Using Au films modified with the DM as electrodes, we fabricated a two-terminal device where a semiconductor layer is bridged over the electrodes, as illustrated in Fig. 2a.[21] The helicene derivative has chirality, which raises the expectation for tunability of the monolayer structure by changing mixing ratios of two optical isomers. A single crystal of rubrene, an archetypal organic semiconductor (OS),[22] was selected as the semiconductor layer to eliminate the possible effects of grain boundaries, which are typically unavoidable in polycrystalline films.

The fabricated devices were characterized using two-terminal electrical measurements, where voltages are applied and the electric current is measured with the two DM-modified Au electrodes. The left panel of Fig. 2b shows the current-voltage (*I-V*) characteristics of the as-fabricated device with an inter-electrode spacing of 0.8 μm and a helicene derivative monolayer with percentage enantiomeric excess (ee) of 75% ee.[21] The relatively uncommon *I-V* curve shape (current increase in the near-zero voltage region followed by quasi-saturation behavior) reflects the double Schottky nature of the device,[23,24] where two metal/semiconductor interfaces both exhibit a charge injection barrier, *i.e.*, are characterized as a Schottky contact.[25] The energy diagram deduced from this is depicted in the left panel of Fig. 2c. A higher voltage of −30 V was applied for 200 s to stimulate a structural change of the DM. The high voltage was applied through the two DM-modified electrodes identical to those used for the low-voltage *I-V* measurements. The central panel of Fig. 2b displays the *I-V* characteristics immediately after high-voltage application; the measured *I-V* curve shape is typical for a diode, where clear rectification behavior is displayed. Although it is presently unclear whether one of two metal/semiconductor interfaces can be characterized as a barrier-less ohmic contact,[26] a significantly asymmetric charge injection barrier height is brought into the two metal contacts.

The energy diagram for this system is depicted in the central panel of Fig. 2c. A higher voltage with opposite polarity, +30 V, was then applied for 200 s; and the low-voltage *I-V* characteristics were measured again and are shown in the right panel of Fig. 2b. The polarity of the diode was reversed, and thus the energy diagram is also reversed as shown in the right panel of Fig. 2c. The polarity of the diode-like characteristics was found to be repeatedly reversed by changing the polarity of the applied high voltage, as shown in Fig. 2d, which indicates that reversible switching of the charge injection barrier is achieved with the two-terminal device with the helicene derivative DM.

The post-modification tunability of metal/OS interfaces has been demonstrated with optically-switchable molecular monolayers.[27] A photochromic molecule is known to change its chemical structure upon light irradiation. Although switching ratios of electric current were only less than two, metallic electrodes modified with a monolayer of a photochromic azobenzene derivative showed change in charge injection efficiency along with the structural change of the derivatives.[27] However, it is noteworthy that the switching based on photochromic molecular layers is achieved with an optic stimulus and not with an electrical stimulus. In actual electric circuits, the semiconductor devices are fully controlled by such electrical signals. The achievement of electrical switching will be significant for molecular electronics. In addition, switching ratios were found to be as high as $10^5$ in the present study, which is sufficiently high for memory application. The rectification behavior was found to survive at least for two days after switching although the rectification ratio (defined by the ratio of the absolute electric currents at ±1 V) decreased from $10^5$ to $10^2$; the devices possess a certain level of non-volatility, and the retention time might become longer by changing the DM molecular structure to increase the molecule/electrode-surface interaction.

The reversible polarity switching of OS-based diodes is reminiscent of resistive switching phenomena that have been reported with various transition metal oxides. Among proposed models for the oxide-based switching, an interface-trap mechanism cannot be completely excluded at present, but it's not very likely for the present case considering the possibly low density of metal/OS interface states.[21] Herein, we propose another possible mechanism where the molecular modification layer plays a significant role: An external electric field interacts with the permanent electric dipoles of the DM molecules, which can lead to a structural change of the DM. The helicene derivative synthesized in this study has two methyl groups (see Fig. 2a), and thus possesses permanent dipoles that point away from the metal surface.[28] An external electric field applies a torque to the dipoles, and the helicene derivative monolayers on the left and right electrodes undergo different structural changes, as illustrated in Fig. 3a. The structural change must affect the magnitude of the two main effects of the DMs on the metal WFs, *i.e.*, the dipole and push-back effects (see Fig. 1).

The magnitude of the dipole effect is dependent on the magnitude of the dipole moment perpendicular to the surface. A change in the metal WF by the dipole effect is expressed as:[5]

$$\Delta \Phi = \frac{qN\mu_\perp}{\varepsilon_0} = \frac{qN\mu_0 \cos(\theta)}{\varepsilon_0 \varepsilon^{\text{eff}}}, \qquad (1)$$

where $q$ is the elementary charge, $N$ is the surface density of DM molecules, $\mu_\perp$ is the normal component of the dipole moment of a single DM molecule, $\varepsilon_0$ is the vacuum permittivity, $\mu_0$ is the dipole moment of an isolated DM molecule, and $\theta$ is the tilt angle of the molecule from the surface normal. $\varepsilon^{\text{eff}}$ represents the effective relative permittivity that takes into account the mutual depolarization of adjacent molecules, which is due to screening of dipole charges by the π-conjugated molecular cores.[2,3] Equation (1) indicates that $\Delta\Phi$ becomes largest for upright standing molecules and smallest for flat-lying molecules on the surface. Thus, for the dipole effect, the expected energy diagrams that correspond to the structural changes depicted in Fig. 3a are constructed in Fig. 3b. However, these energy diagrams are inconsistent with the experimentally deduced diagrams shown in Fig. 2c.

The magnitude of the push-back effect is requisitely dependent on the average distance between the molecular skeletons of the monolayer and the metal surface. The electrons spilled at the metal surfaces are pushed back inside the metal due to Pauli repulsion from the electron clouds of the DM molecular skeletons. The amount of electrons pushed back into the metal increases as the distance decreases. The push-back effect decreases the metal WF by weakening the strength of the surface electric double layer at the metal surface. Expected energy diagrams for the push-back effect are constructed as Fig. 3c (cf. Fig. 1c), and are opposite to that expected for the dipole effect (Fig. 3b), while identical to that deduced from the experimental results (Fig. 2c). We have compared these two effects, and found that the magnitude of the push-back effect can overcome that of the dipole effect in the present system.[21] A similar relationship of the magnitude of the two effects has also been reported for a static work function change of Au(111) upon vanadyl naphthalocyanine adsorption.[29] In these systems, the two effects are oppositely-oriented. A larger switching is expected if the methyl groups of the helicene derivative used in this study are substituted with groups that possess permanent dipoles pointing towards the electrode surface, e.g., with nitro or cyano groups.

In conclusion, a simple strategy that employs a molecule with multiple bonding groups has been proposed to enable switchable characteristics to be imparted to a disordered molecular monolayer formed on a metal electrode surface, where the monolayer is expected to exhibit a structural change in response to an external electric field and this changes the metal WF. Dipole and push-back mechanisms are considered to simultaneously contribute to the change in the metal WF, and the relative magnitudes of the two mechanisms determine the direction of the change. The change in the WF due to the DM-based molecular switch was manifested as a reversible height change of the charge injection barrier formed at the metal/OS contacts. The barrier height is a key parameter to determine the operation of semiconductor devices such as FETs and diodes. Reversible switching of the barrier heights provides the possibility to achieve highly efficient charge injection by fine tuning of the barrier height, and that of the fabrication of functional devices with reconfigurable architectures. In particular, the switching ability would be suitable for the construction of non-volatile memory devices.

This work was partially supported by the World Premier International Research Center Initiative, and Special Coordination Funds for Promoting Science and Technology from the Ministry of Education, Culture, Sports, Science and Technology of Japan; and Advanced Low Carbon Technology Research and Development Program (ALCA) from Japan Science and Technology Agency.

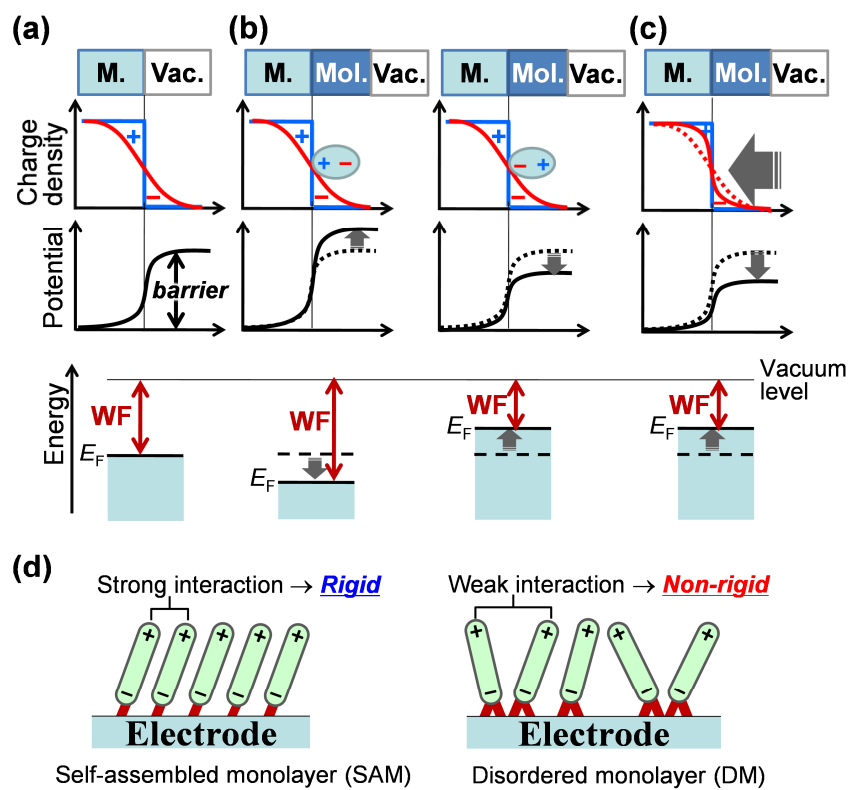

FIG. 1. Two main mechanisms of work function (WF) change by modification of metal surfaces with molecular monolayers. (a) Electric double layer formed at a bare metal surface. Electrons spilling out from the surface create a charge density imbalance, which forms an energy barrier for electrons inside the metal. (b) Change in the WF by the permanent electric dipole of the molecules. The additional electric double layer changes the WF as a function of the strength/direction of the dipole. (c) Change in the WF by the push-back effect. The electrons spilled at the metal surface are pushed back inside the metal due to Pauli repulsion from the electron clouds of the molecules. (d) Schematic of a conventional self-assembled monolayer (left) and a disordered monolayer used in this study (right). The non-rigid DM structure is expected to be controlled by an external electric field, which enables a post-modification change in the WF.

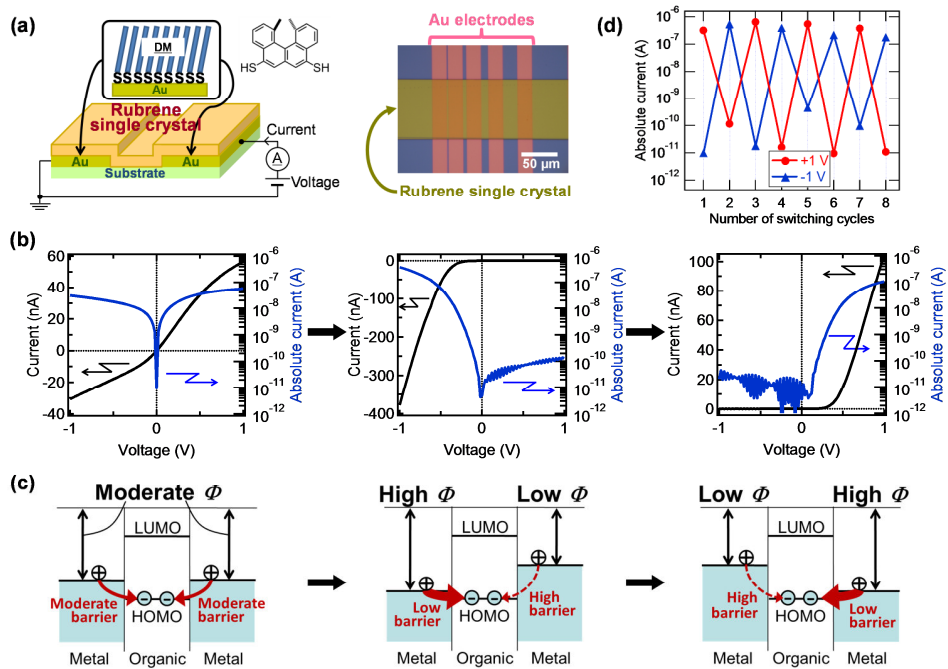

FIG. 2. Reversible switching of charge injection barriers with a DM-modified-metal/organic-semiconductor/DM-modified-metal diode. (a) Schematic (left) and optical micrograph (right) of a constructed device. 1,12-dimethyl-5,8-[4]helicenedithiol was employed as the DM molecule. (b) *I-V* characteristics of a device with a channel length of 0.8 μm and the helicene derivative monolayer at 75% ee.[21] Semi log plots of the absolute current values are also shown in the graphs (blue lines). The device was first measured as-fabricated (left), which revealed double Schottky characteristics. The same device was then measured after application of −30 V for 200 s (center), which exhibited single Schottky characteristics. Finally, the same device was measured again after application of +30 V for 200 s (right), which resulted in a reversal of the polarity of the Schottky diode behavior. Further voltage applications with different polarities were confirmed to reversibly switch the polarity of the diode. (c) Energy diagrams deduced from the *I-V* characteristics measured for the as-fabricated device (left), after application of −30 V (center), and after application of +30 V (right). (d) Repetitive switching of a device with a channel length of 1.8 μm and the helicene derivative monolayer at 0% ee.

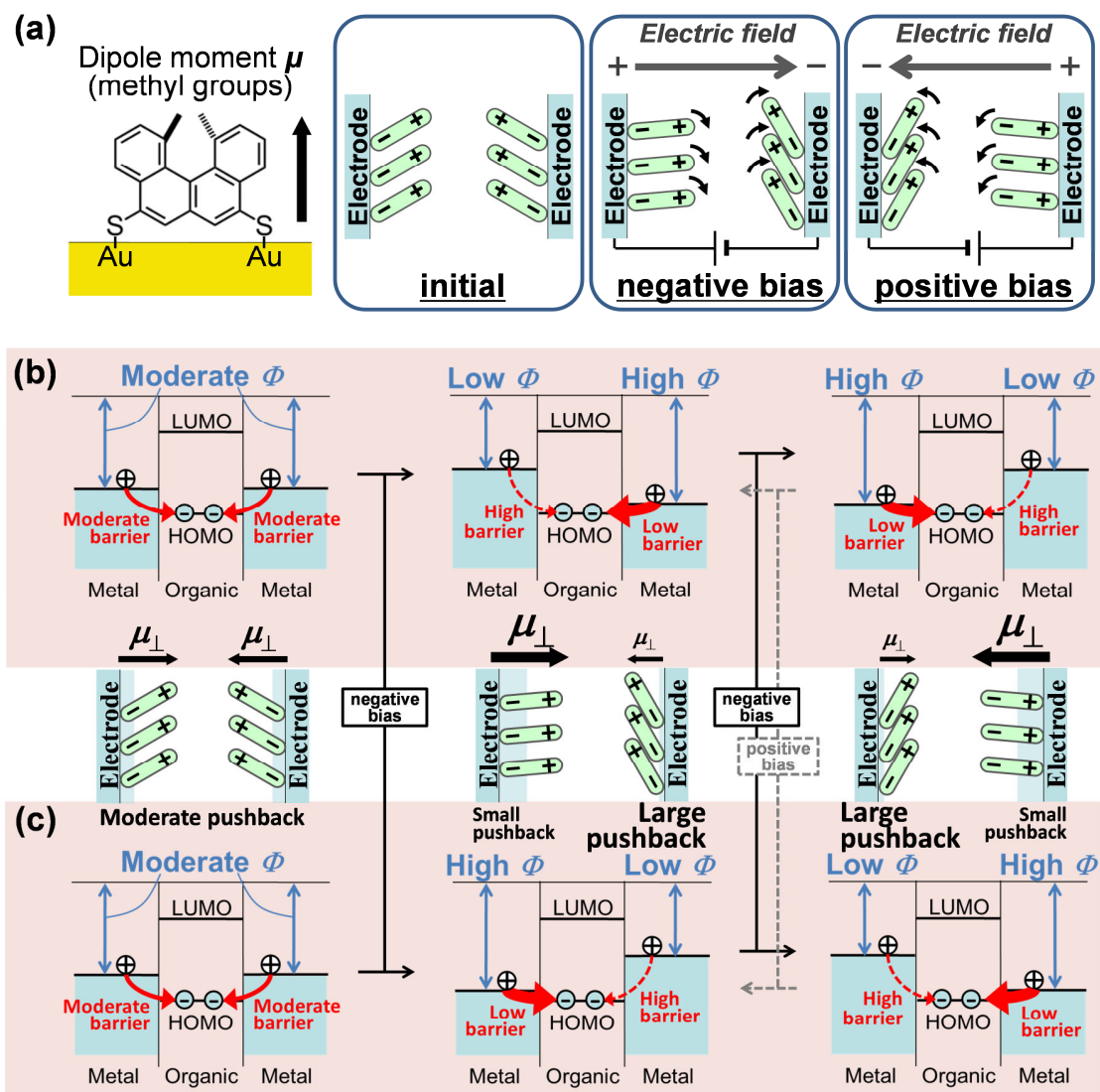

FIG. 3. Mechanism for the reversible switching. (a) Structural switching of DMs due to the permanent electric dipole of the DM molecules. The structural change can be regulated by the direction of an external electric field. (b) Energy diagrams expected for the dipole mechanism, which are opposite to those deduced experimentally (Fig. 2c). The change in the WF is determined by the magnitude of the dipole moment perpendicular to the surface. (c) Energy diagrams expected for the push-back mechanism, which are identical to those deduced experimentally (Fig. 2c). The change in the WF is determined from the distance between the molecular skeletons of the DMs and the metal surface.

Supplementary Material for

# Reversible Switching of Charge Injection Barriers at Metal/Organic-Semiconductor Contacts Modified with Structurally Disordered Molecular Monolayers

Ryo Nouchi,[1,2,3,a)] Masanori Shigeno,[4] Nao Yamada,[3] Tomoaki Nishino,[1] Katsumi Tanigaki[2,3] and Masahiko Yamaguchi[2,4,a)]

[1]N2RC, Osaka Prefecture University, Sakai 599-8570, Japan

[2]WPI-Advanced Institute for Materials Research, Tohoku University, Sendai 980-8577, Japan

[3]Department of Physics, Graduate School of Science, Tohoku University, Sendai 980-8578, Japan

[4]Department of Organic Chemistry, Graduate School of Pharmaceutical Sciences, Tohoku University, Sendai 980-8578, Japan

a) Electronic addresses: r-nouchi@21c.osakafu-u.ac.jp and yama@m.tohoku.ac.jp

1. **Methods**
2. **Choice of solvent for the formation of the helicene derivative monolayer**
3. **Stability against low-voltage application**
4. **Effect of an enantiomeric excess of the helicene derivative on rectification ratios**
5. **Control experiments without the molecular modification layer on electrode**
6. **Mechanisms proposed to explain resistive switching phenomena in transition metal oxides**
7. **Structural analyses of the helicene derivative monolayer on Au surfaces**
8. **Comparison of the magnitudes of the dipole and push-back effects**

1. **Methods**

A heavily doped Si wafer with a thermal oxide layer on top was used as the substrate for device fabrication. Multiple Au electrodes were fabricated on the substrate using electron-beam lithography, vacuum deposition of metal thin films by resistive heating (14 nm thick Au film with a 1 nm thick adhesion layer of Cr), and a subsequent lift-off process. The electrode widths were 1 μm and the inter-electrode spacing was designed to be varied from 0.5 to 100 μm. Prior to formation of the molecular monolayer onto the fabricated Au electrodes, the substrates were cleaned using an oxygen plasma treatment followed by immersion in ethanol for 1 h to reduce the resultant Au oxide layer [S1]. Substrates with Au electrodes were then submerged into a 1 mM tetrahydrofuran solution of the helicene derivative for 24 h to form a molecular monolayer on the electrodes. After monolayer formation, the substrates were rinsed with pure tetrahydrofuran and ultrasonicated for 1 min to desorb physisorbed 1,12-dimethyl-5,8-[4]helicenedithiol molecules on the monolayer and bare SiO2 surface, which left monolayers selectively on the electrode surfaces. Rubrene single crystals were grown using physical vapor transport (PVT) [S2] from a commercially available source powder (90.7% purity, Wako Chemical). The PVT process was repeated three times to improve the purity of the obtained crystals. Freshly prepared crystals were laminated onto the electrodes under ambient conditions immediately after monolayer formation. Measurements of the I-V characteristics and application of high voltages were performed using a semiconductor device analyser (Agilent B1500A) or a sourcemeter (Keithley 2636A). All measurements were conducted at room temperature, in air, and under dark conditions.

2. **Choice of solvent for the formation of the helicene derivative monolayer**

This is the first report on the formation of the helicene derivative monolayer; therefore, the formation conditions of the monolayer were examined first. Ethanol is widely used as a solvent for the formation of self-assembled monolayers of alkanethiols and thiophenols. However, when ethanol was used as a solvent for the helicene derivative, none of the fabricated devices functioned properly (no electric current was detected through the devices). Figure S1(a) shows an optical micrograph of the substrate just after monolayer formation using ethanol. A thick layer on the substrate surface can be discerned from the colour of the electrode surfaces, which is very different from that of the bare surfaces. Thus, ethanol is a poor solvent for the helicene derivative, due to oligomer precipitation via disulphide formation, and the thick oligomer layer blocks electrical conduction from the electrode to the organic semiconductor.

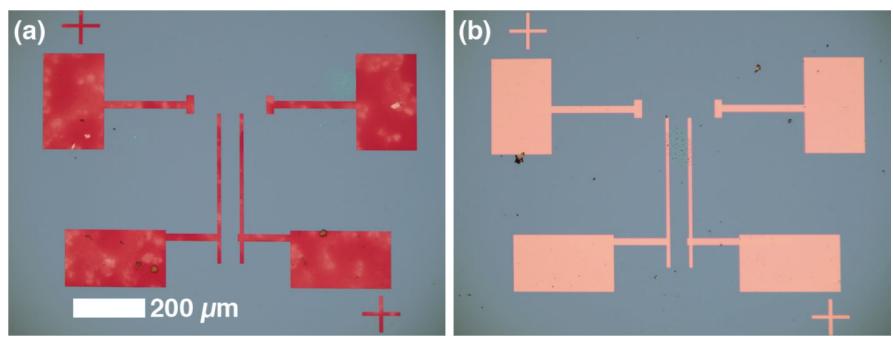

Fig. S1. Optical micrograph of substrates after formation of a helicene derivative layer using (a) ethanol and (b) tetrahydrofuran as solvents.

Tetrahydrofuran (THF) was employed as the next solvent option. No colour change of the electrode surfaces was evident (Fig. S1b), which indicates that THF is a good solvent for the helicene derivative. As expected, most of the devices fabricated using THF as a solvent functioned properly. All of the devices reported in this paper were fabricated using THF.

## 3. Stability against low-voltage application

To characterize the electrical properties, current-voltage ($I$-$V$) characteristics were measured within the low-voltage range (±1 V). As discussed in the main text, application of a higher voltage induces switching of the charge injection barriers. Therefore, to properly characterize the devices, low-voltages, within ±1 V, were applied and the switching characteristics were evaluated. Figure S2 shows the time evolution of $I$ with applied $V$ of −1 and −2 V, which was measured after switching by applying +100 V for 100 s. The inter-electrode spacing of the device was around 1.8 μm in Fig. S2, while that of the device discussed in the main text was 0.8 μm. Thus, the time evolution at −2 V may well explain the stability of the device discussed in the main text. The electrical current at −2 V was found to be stable up to 10 s. Single *I-V* measurements used to characterize devices were completed within 10 s. As a result, low-voltage application within ±1 V does not induce switching.

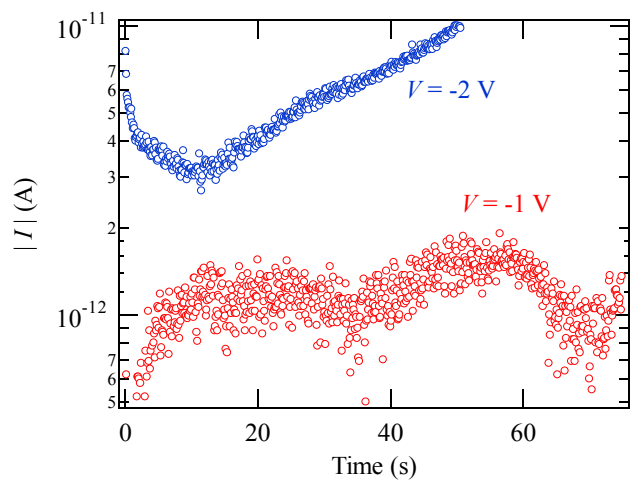

Fig. S2. Time evolution of the electric current during application of a low reverse bias. The channel length of the device was 1.8 μm.

## 4. Effect of an enantiomeric excess of the helicene derivative on rectification ratios

The helicene derivative used in this study is a chiral molecule that has two optical isomers (enantiomers). Thus, a mixture of the two isomers can be prepared with various mixing ratios. The ratio is characterized by a percentage enantiomeric excess (% ee), which is defined by the absolute difference between the mole fraction of each enantiomer. Therefore, 0% ee indicates a 1:1 mixture (a racemic substance), and 100% ee represents optically pure substances. Devices were fabricated using 1 mM THF solution of the helicene derivative with 0, 50, 75, and 100% ee. The number of working devices with 0, 50, 75, and 100% ee were 11, 12, 7, and 8, respectively. Figure S3 shows the percentage yields of devices with a rectification ratio in a certain range. The rectification ratios were determined from *I-V* measurements within the range from −1 V to +1 V after switching the devices at −30 V for 200 s, and were defined by the absolute value of the electric current at −1 V divided by the current at +1 V. Results for devices with various channel lengths from 0.5 to 100 μm are included in the graph.

Figure S3 shows that the rectification ratio does not have a monotonic relationship with the enantiomeric excess. The effects of enantiomeric excess have been reported to be generally monotonic (for example, see refs. S3 and S4). Unlike such general trends, the present result indicates a unique non-monotonic dependence of the rectification ratio on the enantiomeric excess of the helicene derivative monolayer. Although the background mechanism for the non-monotonicity remains to be elucidated, and in addition the numbers of tested devices are limited, the dependence suggests that it might be possible to tune the surface properties beyond the racemic (0% ee) or optically pure (100% ee) limits.

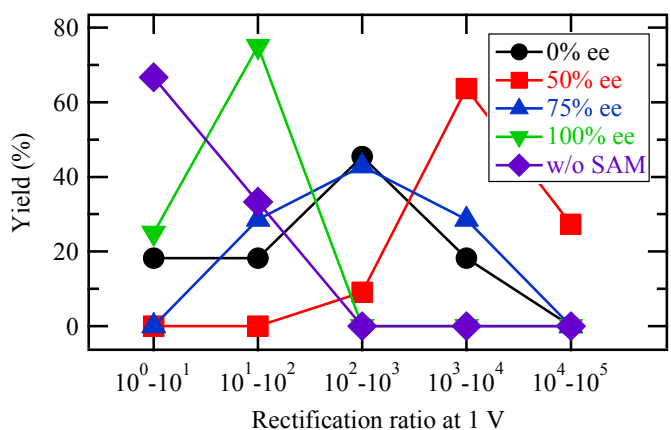

Fig. S3. Percentage yield of devices with rectification ratios in a certain range. The ratio is calculated by dividing the absolute value of the electric current at −1 V by the current at +1 V. Solid lines indicate the data trend.

## 5. Control experiments without the molecular modification layer on electrodes

To further consider the effects of the molecular modification layer on electrode surfaces, control experiments without the modification layer were also carried out. The number of working devices was 6. Figure S3 shows the percentage yields of devices with a rectification ratio in a certain range. The rectification ratios were determined from *I-V* measurements within the range from −1 V to +1 V after switching the devices at −100 V for 140 s, and were defined by the absolute value of the electric current at −1 V divided by the current at +1 V. Results for devices with various channel lengths from 0.8 to 100 μm are included in the graph. Although the number of tested devices is limited, the data trend clearly shows weaker switching strength without the molecular modification layer despite the higher voltage application.

## 6. Mechanisms proposed to explain resistive switching phenomena in transition metal oxides

The reversible polarity switching of organic-semiconductor-based diodes is reminiscent of resistive switching phenomena that have been reported with various transition metal oxides (TMOs). Proposed models for the TMO-based switching can be classified into two types: namely, a conducting filament model and an interface modification model [S5]. In the former model, formation and rupture of conducting filament(s) across the TMO layer defines high resistance and low resistance states, respectively. However, the conducting filament formation is not accountable for the rectification behaviour found in this study. In the latter model, the metal/TMO interface is modified by a redox reaction of electrodes via oxygen exchange with the TMO layer, or by a change in the Schottky barrier height through charge trapping at the interface. In this study, freshly-prepared rubrene single crystals are used, and thus a negligible amount of oxygen is expected in the crystals. Although the charge-trap induced barrier height change with TMOs can be explained mainly by oxygen vacancy redistribution [S6], charge carriers other than the vacancies (electrons and/or holes) should play the same role if there are a certain amount of trap states at the metal contacts. Since rubrene crystals are a van der Waals solid, no dangling bonds exist at their surfaces and a low density of surface states is expected. Metal/OS interface states are known to be induced mostly through metal-molecule coupling [S7]. A greatly reduced density of interface states was reported with weakly-coupled metal contacts to rubrene single crystals [S8]. This situation should hold also in the present case where direct metal-molecule couplings are cut by the electrode surface modification. Actually, control experiments without modification layers showed rather weak switching (Sec. 5 in the supplementary material). In addition, only a mild rectification was reported to be induced to Ti/pentacene/Ti planar devices even with extreme bias conditions of ±500 V at 75°C for 20 h to the device with a channel length of 0.1 mm [S9], while our devices with the same channel length can be effectively switched within 1 min at ±30 V at room temperature. The interface-trap mechanism cannot be completely excluded at present, but it's not very likely for the present case considering the possibly low density of metal/OS interface states.

## 7. Structural analyses of the helicene derivative monolayer on Au surfaces

In the main text, we proposed a mechanism where structural switching of helicene derivative of the modification layer causes a change in the electrode work function. To probe this structural change, we performed Fourier-transform infrared spectroscopy (FTIR) with the surface-sensitive reflection absorption spectroscopy (RAS) configuration (iS50, Thermo Fisher Scientific). The incident angle of the infrared light was set to 75°. Helicene derivative monolayers were prepared on a Au thin film formed on a highly-doped Si substrate with a very thin native oxide layer, where external electric field was applied to the monolayer with the direction normal to the sample surface using lead wires connected to the Si substrate and a conductive plate placed ca. 60 μm away from the surface. If the external electric field causes substantial change in the monolayer structure, the RAS signal changes as well. Since the RAS technique detects signals from transition dipoles normal to the surface, a strong (weak) signal is expected to be obtained from a molecule with a standing (lying) configuration. However, a clear change in the RAS signals was not detected after applying up to 350 V for 5 min through the air gap of ca. 60 μm thick (Fig. S4), though a clear RAS signal can be obtained with the helicene derivative monolayer.

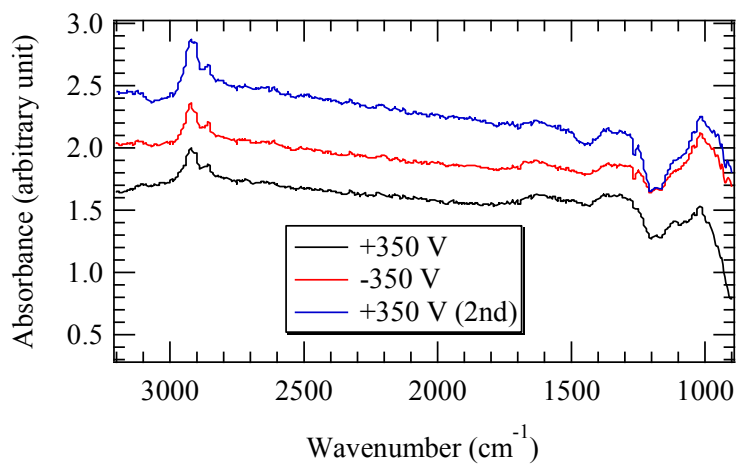

Fig. S4. FTIR-RAS spectra of a helicene derivative monolayer formed on a Au film. The spectra were acquired after each voltage application (+350 V → −350 V → +350 V), and divided by that of the as-fabricated monolayer. The spectra are shifted for clarity. C-H stretching modes (~3000 cm$^{-1}$) and benzene ring derived vibrational modes (< 1700 cm$^{-1}$) are clearly present, but distinct differences between each spectrum are hardly seen.

To detect the difference in RAS signals before and after the external voltage application, the helicene derivative molecule should show a distinct difference in its structure before and after the voltage application. We also conducted a scanning tunnelling microscopic (STM) study on an as-prepared racemic monolayer fabricated on a Au(111) film formed on a freshly-cleaved mica substrate. STM images were obtained in air at room temperature with a sample bias of +750 mV and a constant tunnelling current of 0.45 nA (Nanoscope E, Veeco). A typical molecularly-resolved STM image of the as-prepared helicene derivative monolayer is shown in Fig. S5a. From three topographic scans at different points on the identical sample, the average lateral size of single protrusions was determined to be 1.03 nm (corresponding to the surface density of 9.5×10$^{13}$ cm$^{-2}$). This indicates that most of helicene derivative molecules on a Au surface lie down on the surface (cf. Fig. S5b). In this adsorption configuration, two or three out of four benzene rings constituting the molecular skeleton should make contact with the Au surface. Thus, a change in the adsorption distance of the upper half of the skeleton might be accountable for the switching behaviour found in this study. It is considered to be difficult to detect such a small change in the adsorption configuration by the FTIR-RAS technique; further studies with different molecular species (and/or stronger electric fields) are necessary to detect the structural change of the molecular monolayer by RAS signals.

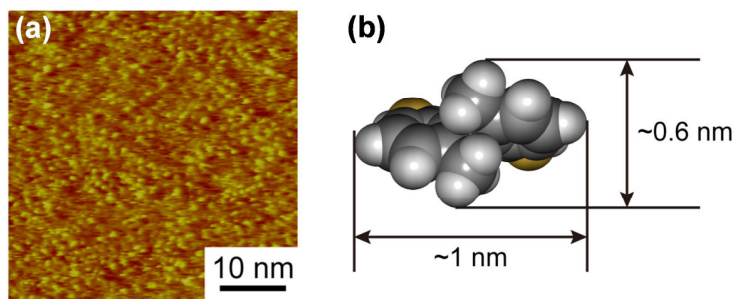

Fig. S5. (a) STM topographic image of a helicene derivative monolayer formed on a Au(111) surface. (b) Molecular structure of a helicene derivative used in this study.

## 8. Comparison of the magnitudes of the dipole and push-back effects

The structural change of disordered monolayers (DMs) can be explained by the applied torque from the external electric field to the permanent electric dipoles of the DM molecules; however, the dipole mechanism does not account for the subsequent change in the metal WF, whereas the push-back mechanism does. This non-intuitive conclusion can be understood by comparing the magnitude of the dipole and push-back effects. A methyl group attached to a benzene ring has a permanent electric dipole of $\mu_0 = 0.37$ D [S10]; $\varepsilon^{\text{eff}}$ of a benzene-based molecule is typically in the range from 2 to 3 [S11]; the surface density of methyl groups, $N$ is $1.9 \times 10^{14}$ cm$^{-2}$ (the double of the helicene derivative density; Sec. 7 in the supplementary material). By substituting these values into Eq. (1) in the main text, $\Delta\Phi$ for the dipole effect is obtained as 0.13 eV at most. In contrast, $\Delta\Phi$ for the push-back effect was reported to be −1.10 eV with flat-lying adsorption of a dense benzene monolayer ($2.4 \times 10^{14}$ cm$^{-2}$) on Au(111) [S12]. From the adsorption configuration revealed by STM measurements (Sec. 7 in the supplementary material), one or two out of four benzene rings constituting the molecular skeleton of the helicene derivative can change their adsorption distance and cause the WF change, and this corresponds to the benzene ring density of $9.5 \times 10^{13}$ or $1.9 \times 10^{14}$ cm$^{-2}$, respectively. Considering $\Delta\Phi$ of −1.10 eV with $2.4 \times 10^{14}$ cm$^{-2}$, these densities lead to $\Delta\Phi$ of at most 0.44 or 0.87 eV, respectively. A simple comparison of $\Delta\Phi$ for the two effects suggests that the magnitude of the push-back effect can overcome that of the dipole effect in the present system.